\documentclass[12pt]{iopart}
\usepackage{iopams}

\usepackage{graphicx} 
\usepackage{subfigure}
\usepackage{epstopdf}
\usepackage{lineno}
\usepackage{multirow}
\usepackage{rotating}

\usepackage{cite,./mcite}
\usepackage{hyperref}

\usepackage{amssymb,amssymb}
\usepackage{amsthm}

\newlength{\figwidth}
\begin{document}
\vspace*{-0.5in}
\begin{flushright}
CERN-PH-EP-2011-030 \\
\end{flushright}

%-------------------------- Macros -----------------------------

\def\integLumi{36~pb$^{-1}$}
\def\integLumiUncertainty{$\pm 11$\%}
\def\JERuncertainty{14} 
\def\MjjCut{200} %
\def\MjjLowerExclusion{400} % 
\def\MjjUpperFitRange{1900} %
\def\MjjLowerLimitMRST{1265} % Note: no units attached!
\def\MjjLowerLimitMRSTEXPECTED{1100} % Note: no units attached!
\def\MjjLowerLimitStatOnlyMRST{1295} % Note: no units attached!
\def\MjjLowerLimitStatOnlyMRSTEXPECTED{1140} % Note: no units attached!
\def\MjjLowerLimitCTEQsixLone{1205} % Note: no units attached!
\def\MjjLowerLimitCTEQsixLoneEXPECTED{1020} % Note: no units attached!
\def\MjjLowerLimitStatOnlyCTEQsixLone{1240} % Note: no units attached!
\def\MjjLowerLimitStatOnlyCTEQsixLoneEXPECTED{xxxx} % Note: no units attached!
\def\MjjLowerLimitCTEQsixsix{1245} % Note: no units attached!
\def\MjjLowerLimitCTEQsixsixEXPECTED{1060} % Note: no units attached!
\def\MjjLowerLimitStatOnlyCTEQsixsix{1280} % Note: no units attached!
\def\MjjLowerLimitStatOnlyCTEQsixsixEXPECTED{xxxx} % Note: no units attached!
\def\MjjLowerLimitMRST{2.15} 
\def\MjjLowerLimitMRSThalfmuRmuF{2.28} 
\def\MjjLowerLimitMRSTtwicemuRmuF{2.05} 
\def\MjjLowerLimitMRSTEXPECTED{2.07} 
\def\MjjLowerLimitStatOnlyMRST{2.27} 
\def\MjjLowerLimitStatOnlyMRSTEXPECTED{2.12} 
%
% CTEQ6L1 (in MC09'[prime] tune)
\def\MjjLowerLimitCTEQsixLone{2.06} % 2060
\def\MjjLowerLimitCTEQsixLoneEXPECTED{2.01} 
\def\MjjLowerLimitStatOnlyCTEQsixLone{2.19} 
\def\MjjLowerLimitStatOnlyCTEQsixLoneEXPECTED{2.07} 
%
% CTEQ5L (in Perugia0 tune)
\def\MjjLowerLimitCTEQfiveL{2.14} 
\def\MjjLowerLimitCTEQfiveLEXPECTED{2.06} 
\def\MjjLowerLimitStatOnlyCTEQfiveL{2.26} 
\def\MjjLowerLimitStatOnlyCTEQfiveLEXPECTED{2.12} 

\def\MjjLowerLimitMRSTAxigluon{2.10} 
\def\MjjLowerLimitMRSTEXPECTEDAxigluon{2.01} 
\def\MjjLowerLimitStatOnlyMRSTAxigluon{2.28} 
\def\MjjLowerLimitStatOnlyMRSTEXPECTEDAxigluon{2.09} 

\def\MjjLowerLimitMRSTQBHntwo{3.20} 
\def\MjjLowerLimitMRSTEXPECTEDQBHntwo{3.18} 
\def\MjjLowerLimitStatOnlyMRSTQBHntwo{3.22} 
\def\MjjLowerLimitStatOnlyMRSTEXPECTEDQBHntwo{3.20} 
\def\MjjLowerLimitMRSTQBHnthree{3.38} 
\def\MjjLowerLimitMRSTEXPECTEDQBHnthree{3.35} 
\def\MjjLowerLimitStatOnlyMRSTQBHnthree{3.39} 
\def\MjjLowerLimitStatOnlyMRSTEXPECTEDQBHnthree{3.37} 
\def\MjjLowerLimitMRSTQBHnfour{3.51}
\def\MjjLowerLimitMRSTEXPECTEDQBHnfour{3.48} 
\def\MjjLowerLimitStatOnlyMRSTQBHnfour{3.52} 
\def\MjjLowerLimitStatOnlyMRSTEXPECTEDQBHnfour{3.50} 
\def\MjjLowerLimitMRSTQBHnfive{3.60} 
\def\MjjLowerLimitMRSTEXPECTEDQBHnfive{3.58} 
\def\MjjLowerLimitStatOnlyMRSTQBHnfive{3.61} 
\def\MjjLowerLimitStatOnlyMRSTEXPECTEDQBHnfive{3.59} 
\def\MjjLowerLimitMRSTQBHnsix{3.67} 
\def\MjjLowerLimitMRSTEXPECTEDQBHnsix{3.64} 
\def\MjjLowerLimitStatOnlyMRSTQBHnsix{3.68} 
\def\MjjLowerLimitStatOnlyMRSTEXPECTEDQBHnsix{3.66} 

\def\MjjLowerLimitMRSTQBH{3.67} 
\def\MjjLowerLimitMRSTEXPECTEDQBH{3.64} 
\def\MjjLowerLimitStatOnlyMRSTQBH{3.68} 
\def\MjjLowerLimitStatOnlyMRSTEXPECTEDQBH{3.66} 
\def\MjjLowerLimitMRSTQBHnseven{3.73} 
\def\MjjLowerLimitMRSTEXPECTEDQBHnseven{3.71} 
\def\MjjLowerLimitStatOnlyMRSTQBHnseven{3.74} 
\def\MjjLowerLimitStatOnlyMRSTEXPECTEDQBHnseven{3.72} 
\def\FchimjjLambda{9.5}  
\def\FchimjjEXPECTEDLambda{5.7}  
\def\FchimjjKFacQCDCILambda{8.1}  
\def\FchimjjKFacQCDCIEXPECTEDLambda{5.9}  
\def\FchiKFacQCDCILambda{6.8}  
\def\FchiKFacQCDCIEXPECTEDLambda{5.2} 
\def\FchimjjQstar{2.64}
\def\FchimjjEXPECTEDQstar{2.12}
\def\FchimjjQBH{3.78}
\def\FchimjjEXPECTEDQBH{3.49}
\def\FchiQBH{3.69}
\def\FchiEXPECTEDQBH{3.37}
\def\FchiLowerLimitMRSTQBHntwo{3.26} 
\def\FchiLowerLimitMRSTEXPECTEDQBHntwo{2.91}
\def\FchiLowerLimitMRSTQBHnthree{3.41}
\def\FchiLowerLimitMRSTEXPECTEDQBHnthree{3.08}
\def\FchiLowerLimitMRSTQBHnfour{3.53} 
\def\FchiLowerLimitMRSTEXPECTEDQBHnfour{3.20} 
\def\FchiLowerLimitMRSTQBHnfive{3.62}
\def\FchiLowerLimitMRSTEXPECTEDQBHnfive{3.29}
\def\FchiLowerLimitMRSTQBHnsix{3.69}
\def\FchiLowerLimitMRSTEXPECTEDQBHnsix{3.37}
\def\FchiLowerLimitMRSTQBHnseven{3.75}
\def\FchiLowerLimitMRSTEXPECTEDQBHnseven{3.43}

\def\ElevenBinChiEXPECTEDLambda{5.4}
\def\ElevenBinChiLambda{6.6}
\def\ElevenBinChiEXPECTEDQBH{3.36}
\def\ElevenBinChiQBH{3.49}

\def\MjjLowerExclusion{0.60}
\def\HighestDijetMass{3.5}

\def\NpseudoExperiments{$10^3$}
\def\NpseudoExperimentsQCDfit{$10^4$}
\def\triggerThreshold{15} % Note: no units attached!
\def\pTcut{80} % Note: no units attached!
\def\topbottomLabel{~}
\def\leftrightLabel{~}
\def\Pythia{{\sc Pythia}}
\def\BlackMax{{\sc BlackMax}}
\def\Geant{{\sc Geant4}}
\def\CalcHEP{{\sc CalcHEP}}
\def\NLOJET{{\sc NLOJET++}}
\def\BumpHunter{{\sc BumpHunter}}
\def\GeVcc{${\rm GeV}$}
\def\TeVcc{${\rm TeV}$}
\def\GeVc{${\rm GeV}$}
\def\invpb{${\rm pb}^{-1}$}
\def\mFchi{F_{\chi}}
\def\mFchimjj{F_{\chi}(m_{jj})}
\def\Fchi{$\mFchi$}
\def\Fchimjj{$\mFchimjj$}
\def\mjj{m_{jj}}
\def\qstar{{q^\ast}}

\def\pT{$p_{\rm T}$}
\def\mpT{p_{\rm T}}
\def\Acc{{\cal A}}   
\def\thetanp{\theta_{np}}   

%-------------------------- Start of Document -----------------------------

\title[Search for New Physics in Dijet Distributions with the ATLAS Detector]{Search for New Physics in Dijet Mass and Angular Distributions in 
$pp$ Collisions at $\sqrt{s}=7$~TeV Measured with the ATLAS Detector}

\author{The ATLAS Collaboration}

\submitto{\NJP}

\date{18 March 2011}

%----------------------------- Abstract -----------------------------

\begin{abstract}
A search for new interactions and resonances produced in 
LHC proton-proton ($pp$) collisions at a centre-of-mass
energy $\sqrt{s}=7$~TeV has been performed with the ATLAS detector.
Using a data set with an integrated luminosity of \integLumi, 
dijet mass and angular distributions  have been measured up to
dijet masses of $\sim \HighestDijetMass$~TeV and found to be in good agreement with 
Standard Model predictions.  
This analysis  sets limits at 95\%\ C.L. on various models for new physics:  an excited quark is excluded with mass between 0.60 and \FchimjjQstar~TeV, an axigluon hypothesis is excluded for axigluon masses between 0.60 and \MjjLowerLimitMRSTAxigluon~TeV and Randall-Meade quantum black holes are excluded in models with six extra space-time dimensions for quantum gravity scales between 0.75 and \MjjLowerLimitMRSTQBH~TeV.  
Production cross section limits as a function of dijet mass are set using a simplified Gaussian signal model to facilitate comparisons with other hypotheses.
Analysis of the dijet angular distribution using a 
novel technique simultaneously employing the dijet mass excludes
quark contact interactions with a compositeness scale $\Lambda$ below \FchimjjLambda~TeV.
\end{abstract}


\maketitle

%------------------------------ Body of Paper -------------------------

\section{Introduction}

The search for new phenomena in particle interactions is perhaps most exciting when new vistas are opened up by significant increases in experimental sensitivity, either by collecting larger samples of data or entering kinematic regimes never before explored.
Searches are particularly compelling when one can do both, as has recently become the case in the first studies of $pp$\ collisions at a centre-of-mass (CM) energy of 7~TeV produced at the CERN Large Hadron Collider (LHC).  
We report on a search for  massive objects and new interactions using a sample of \integLumi\ of integrated luminosity observed by the ATLAS detector.

This analysis focuses on those final states where two very energetic jets of particles are produced with large transverse momentum (\pT) transfer.
These 2~$\to$~2
scattering processes are well described within the Standard Model (SM)  by perturbative
quantum chromodynamics (QCD), the quantum field theory of strong interactions.  
However, there could be additional contributions from the production of a new massive particle that then decays into a dijet final state, or the rate could be enhanced through a new force that only manifests itself at very large CM energies.

One can perform sensitive searches for new phenomena by studying both the dijet invariant mass, $\mjj$, and the angular distributions of energetic jets relative to the beam axis, usually described by the polar scattering angle in the two-parton
CM frame, $\theta^{*}$.
QCD calculations predict that high-\pT\ dijet
production is dominated by $t$-channel gluon exchange, leading to rapidly falling $\mjj$\ distributions and angular distributions
that are peaked at $|\mathrm{cos}\theta^{*}|$ close to 1.
By contrast, models of new processes characteristically predict angular
distributions that would be more isotropic than those of QCD.
Discrepancies from the predicted QCD behaviour would be evidence for new physics.
This analysis focuses on a study of dijet mass and angular distributions, which have been shown by previous studies \cite{Arnison1986244,Bagnaia1984283,CDF:2009DijetSearch,DZero:2009DijetAng,ATLAS:2010bc,ATLAS:2010eza,CMS:2010dijetmass,CMS:2010centrality,CMS:2011DijetMassAngle}\
to be sensitive to new
processes. 
These dijet variables are well suited for searches employing early LHC data.  
The dijet mass analyses can be performed using data-driven background estimates, while the angular analyses can be designed to have reduced sensitivity to 
the systematic uncertainties associated with the jet energy
scale (JES) and integrated luminosity. 

Following on the first ATLAS studies of massive dijet events with 0.3~\invpb\ \cite{ATLAS:2010bc}\ and 3.1~\invpb\ \cite{ATLAS:2010eza}, the full 2010 data set has increased statistical power by more than an order of magnitude, and we have made several improvements to the analysis.  
A variety of models of new physics have been tested and the angular distributions have been analyzed using a new
technique that finely bins the data in dijet mass to maximise the sensitivity of the search to both resonant and non-resonant phenomena.  
We set limits on a number of models and provide cross section limits using a simplified Gaussian signal model to facilitate tests of other hypotheses that we have not considered.

Section~\ref{sec:kinematics}\ describes the kinematic variables we used in this search.  
Section~\ref{sec:selection}\ describes the detector and the data sample, as well as the common event selection criteria used for the studies reported here.
Section~\ref{sec:theorymodels}\ describes the theoretical models employed, including the procedures used to account for detector effects.
Section~\ref{sec:Mjj}\ describes the search for resonance and threshold phenomena using the dijet invariant mass.
Section~\ref{sec:Ang}\ describes the studies employing the angular distributions as a function of the invariant mass of the dijet system.
Section~\ref{sec:conclusions}\ summarises our results.

\section{Kinematics and Angular Distributions}
\label{sec:kinematics}

This analysis is focused on those $pp$\ collisions that produce two high energy jets recoiling back-to-back in the partonic CM frame to conserve momentum relative to the beamline.
The dijet invariant mass, $\mjj$, is defined as the mass of the two highest \pT\ jets in the event.
The scattering angle $\theta^{*}$\ distribution for $2\to2$ parton scattering is predicted by
QCD in the parton CM frame, which is in practice moving 
along the beamline due to the different  momentum fraction (Bjorken $x$)
of one incoming parton relative to the other.
The rapidity of each jet is therefore a natural variable for the study of these systems, 
$y \equiv \frac{1}{2}\mathrm{ln}(\frac{E + p_z}{E - p_z})$,
where $E$\ is the jet energy and $p_z$\ is the $z$-component of the jet's momentum
\footnote{
The ATLAS coordinate system is a right-handed Cartesian
system with the $x$-axis pointing to the centre of the LHC ring, the $z$-axis
following the counter-clockwise beam direction, and the $y$-axis directed upwards.
The polar angle $\theta$ is referred to the $z$-axis, and $\phi$ is the azimuthal
angle about the $z$-axis.
Pseudorapidity is defined as $\eta \equiv -\ln\tan\left(\theta/2\right)$\ and is a good approximation to rapidity as the particle mass approaches zero.
}.
The variable $y$ transforms under a Lorentz boost along the $z$-direction as
$y \to y - y_B = y - \mathrm{tanh}^{-1}(\beta_B)$,
where $\beta_B$ is the velocity of the boosted frame,
and $y_B$ is its rapidity boost.

We use the $\mjj$\ spectrum as a primary tool in searching for new particles that would be
observed as resonances.  
The $\mjj$\ spectrum is also sensitive to other phenomena, such as threshold enhancements or the onset of new interactions at multi-TeV mass scales in our current data sample.
We bin the data in $\mjj$\ choosing bin-widths that are consistent with the detector resolution as a function of mass so that binning effects do not limit our search sensitivity.
 
We employ the dijet angular
variable $\chi$\ derived from the rapidities of the two highest \pT\ jets, $y_1$\ and $y_2$.
For a given scattering angle $\theta^*$, the corresponding rapidity in the parton CM frame
(in the massless particle limit) is
$y^* = \frac{1}{2}\mathrm{ln}(\frac{1 + |\mathrm{cos}\theta^*|}{1 - |\mathrm{cos}\theta^*|})$.
We determine $y^*$ and $y_B$ from the rapidities of the two jets using
$y^* = \frac{1}{2}(y_1 - y_2)$ and $y_B = \frac{1}{2}(y_1 + y_2)$.
The variable $y^*$\ is used to determine the partonic CM angle $\theta^*$\ and to define
$\chi \equiv \mathrm{exp}(|y_1-y_2|)=\mathrm{exp}(2|y^*|)$.
As noted in previous studies, the
utility of the $\chi$ variable arises because the $\chi$\ distributions associated with final states produced via QCD interactions are relatively flat compared with the distributions associated with new particles or interactions that typically peak at low values of $\chi$.

In a previous dijet angular distributions analysis~\cite{ATLAS:2010eza},
a single measure of isotropy based on $y^*$\ intervals was introduced.
This measure, \Fchi, is the fraction of dijets produced centrally versus the 
total number of observed dijets for a specified dijet mass range.  We
extend this to a measure that is finely binned in dijet mass intervals:
\begin{eqnarray}
&&    F_{\chi}\left(\left[m^{min}_{jj}+m^{max}_{jj}\right]/2\right) \equiv  \nonumber \\
&& \qquad\qquad\qquad  \frac{N_{events}(|y^*|<0.6,m^{min}_{jj},m^{max}_{jj})}{N_{events}(|y^*|<1.7,m^{min}_{jj},m^{max}_{jj})},
\label{eq:vapprox}
\end{eqnarray}
where $N_{events}$\ is the number of candidate events within the $y^*$\ interval and in the
specified $\mjj$\ range.
The interval $|y^*|<0.6$\ defines the central region where we expect to be most 
sensitive to new physics and corresponds to the angular region $\chi < 3.32$, while
$|y^*|<1.7$\ extends the angular range to $\chi < 30.0$, where QCD processes dominate.
This new observable, \Fchimjj,
is defined using the same fine $\mjj$\ binning used for analysis of the $\mjj$\ spectrum.
We also employ the variable \Fchi\ to denote the ratio in Eq.~\ref{eq:vapprox}\ for dijet masses above 2~TeV.
Our studies have shown that the \Fchimjj\ distribution is sensitive to mass-dependent changes in the rate of
centrally produced dijets. 

Jets are reconstructed using the infrared-safe anti-$k_t$ jet clustering 
algorithm~\cite{antikT,Cacciari:2006} with the distance parameter $R = 0.6$.
The inputs to this algorithm are clusters of calorimeter cells defined by energy depositions significantly
above the measured noise.  
Jet four-momenta are constructed by the vectorial addition of
cell clusters, treating each cluster as an ($E$, $\vec{p}$) four-vector with zero mass. 
The jet four-momenta are then corrected as a function of $\eta$\ and \pT\ for various effects, the largest of which are the hadronic shower response and detector material distributions. 
This is done using a calibration scheme
based on Monte Carlo (MC) studies including full detector simulation, and validated 
with extensive test-beam studies~\cite{Adragna:2010zz} and collision data \cite{JES,JES2,ATLAS:JES3}.

The measured distributions include corrections for the jet energy scale but are not unfolded to account for resolution effects.  
These distributions are compared to theoretical predictions processed through a full detector simulation software.

\section{ The ATLAS Detector and Event Selection}
\label{sec:selection}
\subsection{The Detector and Trigger Requirements}

The ATLAS detector~\cite{DetectorPaper} is instrumented over almost the entire solid angle around 
the $pp$\ collision point with layers of tracking detectors, calorimeters, and muon chambers.
Jet measurements are made using a finely segmented calorimeter system designed to efficiently detect the high energy jets that are the focus of our study.

The electromagnetic (EM) calorimeter consists of an accordion-shaped lead absorber over the region  $|\eta|< 3.2$, using
liquid argon (LAr) as the active medium to measure the energy and geometry of the showers arising from jets.
The measurement of hadronic energy flow in the range $|\eta| < 1.7$\ is complemented by a sampling calorimeter 
made of steel and scintillating tiles. 
In the end-cap region $1.5 < |\eta| < 3.2$, hadronic calorimeters consisting of steel absorber and a LAr active medium
match the outer $|\eta|$ limits of the EM calorimeters. 
To complete the $\eta$ coverage to $|\eta|<4.9$, the LAr forward calorimeters provide both
EM and hadronic energy measurements.
The calorimeter ($\eta, \phi$) granularities are $\sim0.1 \times 0.1$ for the hadronic
calorimeters up to $|\eta| < 2.5$\
and then $0.2 \times 0.2$ up to $|\eta| < 4.9$. 
The EM calorimeters feature a finer readout granularity varying
by layer, with cells as small as $0.025 \times 0.025$ extending over $|\eta| < 2.5$.

The inner tracking detector (ID) covers the range $|\eta| < 2.5$, and
consists of a silicon pixel detector, a silicon microstrip detector (SCT) and, for
$|\eta| < 2.0$, a transition radiation tracker (TRT). The ID is surrounded by a thin
superconducting solenoid providing a 2T magnetic field.

ATLAS has a three-level trigger system, with the first level trigger (L1) being custom-built hardware and
the two higher level triggers (HLT) being realised in software.
The triggers employed for this study selected events that had at least one large transverse energy deposition, with the transverse energy threshold varying over the period of the data-taking as the instantaneous luminosity of the LHC $pp$\ collisions rose.

\begin{figure}[t]
  \centering
              {
      \includegraphics[width=0.60\textwidth]{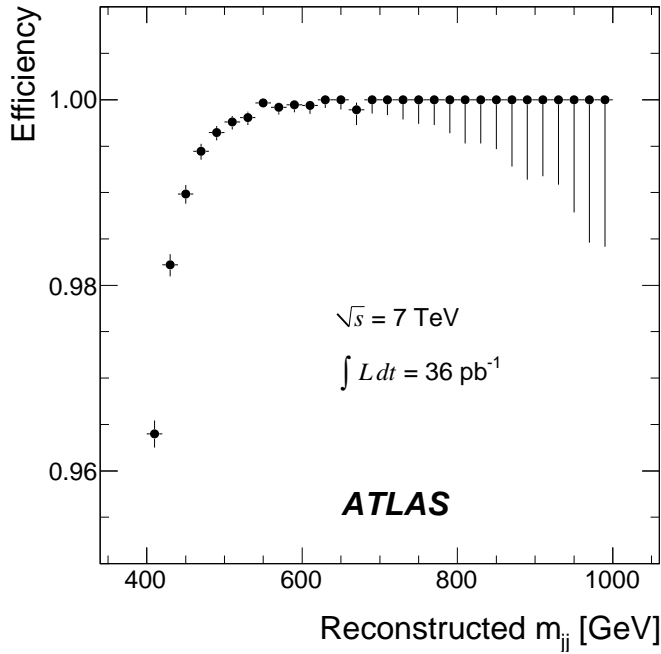}
                    }
    \caption{The efficiency for passing the primary first-level trigger as a function
    of the dijet invariant mass, $\mjj$.
    The uncertainties are statistical.
    }
\label{fig:triggers}
\end{figure}

The primary first-level jet trigger used in the resonance analysis
had an efficiency $>99$\%\ for events with dijet masses 
$\mjj > 500$~GeV. This is illustrated
in Fig.~\ref{fig:triggers}\ where we show the measured trigger efficiency
as a function of $\mjj$.  After applying the full event
selection from the resonance analysis (except the $\mjj$ cut)
we compute the fraction of events passing a reference trigger
which also pass our analysis trigger.  The reference trigger
is an inclusive jet trigger that was fully efficient for
$\mpT> 80$~GeV, while our event selection already requires $\mpT > 150$~GeV
to guarantee full efficiency of the reference trigger.  Thus,
we efficiently identify events for the dijet resonance analysis
for $\mjj > 500$~GeV.

In order to have uniform acceptance for the angular distribution analysis, additional lower-\pT\ triggers were used for different angular and mass regions.
We verified that these triggers provided uniform acceptance as a function of $\chi$\ for the dijet mass intervals in which they were employed.
Because these lower threshold triggers sampled only a subset of the $pp$\ collisions at higher instantaneous luminosity, the effective integrated luminosity collected for  dijet masses between 500 and 800 GeV was 2.2~\invpb\ and between 800 and 1200~GeV was 9.6~\invpb\ in the dijet angular distribution analysis. Above 1200 GeV the same trigger is used for the resonance and angular
analyses, and the full 36~\invpb\ are used for both analyses.

\subsection{Common Event Selection}
Events are required to have at least one primary collision vertex defined by more than four charged-particle tracks.
Events with at least two jets are retained if the highest \pT\ jet
(the `leading' jet) satisfies $\mpT^{{j_1}} > 60$~GeV and the next-to-leading
jet satisfies $\mpT^{j_2} > 30$~GeV. 
The asymmetric thresholds avoid suppression of events where
a third jet has been radiated, while the 30~GeV threshold ensures that reconstruction
is fully efficient for both leading jets.  
Events containing a
poorly measured jet \cite{JetEtMiss_Cleaning_Note} with $\mpT > 15$~GeV are vetoed to avoid
cases where such a jet would cause incorrect identification of the two leading jets.
This criterion rejects less than 0.6\%\ of the events.
The two leading jets are required to satisfy quality criteria that ensure that they arise from in-time energy deposition.

Further requirements are made on the jets in order to optimise the analysis of the dijet mass spectrum and the study of the dijet angular distributions, described in Sections~\ref{sec:Mjj}\ and \ref{sec:Ang}, respectively.

\section{Theoretical Models and Monte Carlo Simulations}
\label{sec:theorymodels}

The MC signal samples used for the analysis have been produced with a variety of event generators.
We have employed several of the most recent parton distribution functions (PDFs) so that 
we consistently match the orders of the matrix element calculations implemented in
the different MC generators when we calculate
QCD predictions, and to be conservative in the calculation of expected new physics signals 
(all new physics signals are calculated only to leading order).  
 
 \subsection{QCD Production}
The angular distribution analyses required a prediction for the angular distribution arising
from QCD production.
Monte Carlo samples modelling QCD dijet production were created with
the \Pythia\ 6.4.21 event generator \cite{pythia}\ and the ATLAS MC09 parameter tune 
\cite {ATLAS_MC09}, using the modified leading-order MRST2007 \cite{Sherstnev:2007nd}
PDF (MRST2007LO*).
The generated events were passed through the detailed simulation of the ATLAS detector
\cite{ATLSIM}, which uses the \Geant\ package~\cite{Agostinelli:2002hh} for simulation of particle 
transport, interactions, and decays, to incorporate detector effects.
The simulated events were then reconstructed in the same way 
as the data to produce predicted dijet mass and angular distributions that can be compared with the observed distributions.

Bin-by-bin correction factors (K-factors) have been applied
to the angular distributions derived from MC calculations to account
for next-to-leading order (NLO) contributions. 
These K-factors were derived from dedicated MC samples and are defined as
the ratio $NLO_{ME}$/$PYT_{SHOW}$.  
The $NLO_{ME}$ sample was produced 
using matrix elements in NLOJET++ \cite{Nagy1,Nagy2,catani-1998-510} with the NLO
PDF from CTEQ6.6 \cite{Nadolsky:2008zw}. 
The $PYT_{SHOW}$ sample
was produced with the \Pythia\ generator restricted to leading-order (LO) matrix elements
and parton showering using the MRST2007LO* PDF.

The angular distributions generated with the full \Pythia\ calculation include
various non-perturbative effects
such as multiple parton interactions and hadronization.
The K-factors defined above were designed to retain these effects while
adjusting for differences in the treatment of perturbative effects.
We multiplied the full \Pythia\ predictions of angular distributions by these
bin-wise K-factors to obtain a reshaped spectrum that includes corrections
originating from NLO matrix elements.  
Over the full range of $\chi$, the K-factors change the normalised angular 
distributions by up to 6\%, with little variability from one mass bin to the other. 

The QCD predictions used for comparison to the measured angular distributions in this article are
the product of the two-step procedure described above.

\subsection{Models for New Physics Phenomena}
MC signal events for a benchmark beyond-the-Standard-Model resonant process 
were generated using the excited-quark
($qg\to \qstar$) production model~\cite{Baur:1987ga,Baur:1989kv}.
The excited quark $\qstar$ was assumed to have spin 1/2 and quark-like
couplings, relative to those of the SM $SU(2)$, $U(1)$, and $SU(3)$
gauge groups, of $f=f^\prime=f_s=1$, respectively.  The compositeness
scale ($\Lambda$) was set to the $\qstar$\ mass.  
Signal events were produced using the \Pythia\ event generator
%, a leading-order parton-shower MC generator, 
with the MRST2007LO* PDF
and with the renormalization and factorization scales set to the mean
\pT\ of the two leading jets. 
We also used the \Pythia\ MC generator  to decay
the excited quarks to all possible SM final states, which are
dominantly $qg$ but also $qW$, $qZ$, and $q\gamma$.  
The MC samples were produced using the ATLAS MC09 parameter tune.

We also considered two other models of new physics that generate resonant signatures:  
axigluons and Randall-Sundrum (RS) gravitons.
The axigluon interaction~\cite{Frampton1987a,Frampton1987b,Bagger1988}\ is described by the Lagrangian
\begin{eqnarray}
\mathcal{L}_{Aq\bar{q}} = g_{QCD}\bar{q}A^{a}_{\mu}\frac{\lambda^a}{2} \gamma^{\mu}\gamma_{5}\ q.
\end{eqnarray}
The parton-level events were generated using the \CalcHEP\ Monte Carlo package~\cite{Pukhov:2004ca}\
with the MRST2007LO* PDF.
We used a \Pythia\ MC calculation to model the production and decays of an RS graviton\cite{Randall:1999ee,Bijnens:2001gh}\ of a given mass. 
We performed this calculation with the dimensionless coupling $\kappa/\bar{M}_{Pl}=0.1$,
where $\bar{M}_{Pl}$ is the reduced Planck mass, to set
limits comparable to other searches \cite{CDF:2007RSGDiphoton,CMS:2010dijetmass}.

For non-resonant new phenomena, we used a benchmark quark contact interaction
as the beyond-the-Standard-Model process.
This  models the onset of kinematic properties that
 characterise quark compositeness:
the hypothesis that quarks are composed of more fundamental particles.
The model Lagrangian is a four-fermion contact interaction 
\cite{Eichten:1984eu,Eichten:1995akc,Chiappetta1991}\ whose
effect  appears below or near a characteristic energy scale $\Lambda$.  
While a number of contact terms are possible, the Lagrangian in standard use since
1984 \cite{Eichten:1984eu} is the single (isoscalar) term: 
\begin{eqnarray}
\mathcal{L}_{qqqq}(\Lambda)=\frac{{\xi}g^2}{2\Lambda^2}
\bar{\Psi}^L_q\gamma^{\mu}\Psi^L_q\bar{\Psi}^L_q\gamma_{\mu}\Psi^L_q,
\end{eqnarray}
where $g^2/4\pi=1$\ and the quark fields
$\Psi^L_q$ are left-handed. 
The full Lagrangian used for hypothesis testing is
then the sum of $\mathcal{L}_{qqqq}(\Lambda)$ and the QCD Lagrangian.  
The relative phase of these terms is controlled by the interference parameter,
$\xi$, which is set for destructive interference ($\xi = +1$) in the current analysis.
Previous analyses \cite{CDF:2009DijetSearch}\ showed that the choice of 
constructive ($\xi = -1$) or destructive ($\xi = +1$) interference changed exclusion 
limits by $\sim 1\%$.
MC samples were created by a \Pythia\ 6.4.21 calculation using this Lagrangian, with each
sample corresponding to a distinct value of $\Lambda$.  

As another example for non-resonant new physics phenomena, we considered Randall-Meade quantum black holes (QBH) \cite{RandallMeade}. 
We used the \BlackMax\ black hole event generator \cite{Dai:2007bm}\ to simulate the 
simplest two-body final state
scenario describing the production and decay of a Randall-Meade
QBH for a given fundamental quantum gravity scale $M_D$.
These would appear as a threshold effect that also depends on the number of extra space-time dimensions.

Previous ATLAS jet studies \cite{jetProdICHEP} have shown that the use of 
different event generators and models for non-perturbative behaviour has a negligible effect on
the observables in the kinematic region we are studying.
All of the MC signal events were modelled with the full ATLAS detector simulation.

% ------------------------------------  Dijet Mass Distribution Search ----------------------------------------%

\section{Search for Dijet Resonances}
\label{sec:Mjj}

We make a number of additional selection requirements on the candidate events to optimise
the search for effects in the dijet mass distribution.
Each event is required to have its two highest-\pT\ jets satisfy   
$|\eta_j| < 2.5$\  with  $|\Delta\eta_{jj}|< 1.3$.
In addition,  the leading jet
must satisfy $\mpT^{j_1} > 150$~GeV and $\mjj$\ must be greater than 500 GeV.
These criteria have been shown, based on studies of expected signals and QCD background, to
efficiently optimise the signal-to-background in the sample.
There are 98,651 events meeting these criteria.

\subsection{The Dijet Mass Distribution}

\begin{figure}
  \centering \includegraphics[width=0.60\textwidth]{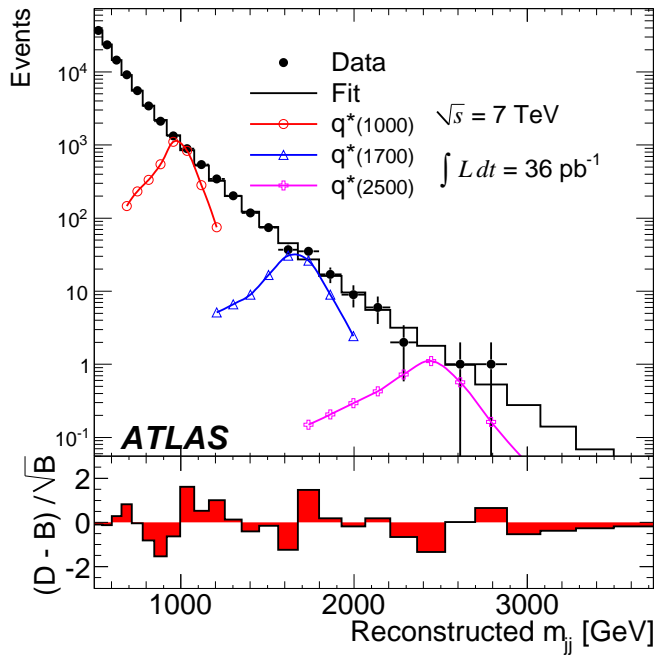}
  \caption{The observed (D) dijet mass distribution (filled points) fitted using a binned QCD background (B) distribution described by Eqn.~\ref{eq:f} (histogram).  The predicted $\qstar$\ signals normalised to \integLumi\ for excited-quark masses of 1000, 1700, and 2500~\GeVcc\ are overlaid.
The bin-by-bin significance of the data-background difference is shown in the lower panel.}
  \label{fig:1}
\end{figure}

In order to develop a data-driven model of the QCD background shape, 
a smooth functional form
	\begin{eqnarray}
	f(x) = p_1 (1 - x)^{p_2} x^{p_3 + p_4\ln x},
	\label{eq:f}
	\end{eqnarray}
where  $x \equiv \mjj/\sqrt{s}$\ and the $p_i$\ are fit parameters, 
is fit to the dijet mass spectrum.
Although not inspired by a theory, this functional form has been empirically shown to model the steeply falling QCD dijet 
mass spectrum~\cite{CDF:2009DijetSearch,ATLAS:2010bc,CMS:2010dijetmass}.
Figure \ref{fig:1} shows the resulting mass spectrum and fitted background, indicating that the observed spectrum is consistent with a rapidly falling, smooth distribution.
The bin widths have been chosen to be consistent with the dijet mass resolution, increasing from 
$\sim50$\ to $\sim200$~GeV for dijet masses from 600 to 3500~GeV, respectively.
The p-value of the fit to the data, calculated using the chi-squared determined from
pseudo-experiments as a goodness-of-fit statistic,  is 0.88.
Although this p-value suggests that there is no significant overall disagreement, we use a more sensitive statistical test, the \BumpHunter\ algorithm~\cite{Aaltonen:2008vt,Choudalakis:2011bh}, to establish the presence or absence of a resonance.

In its implementation for this analysis, the \BumpHunter\ algorithm  searches for the signal window with the most significant excess of events above the background, requiring insignificant discrepancy (Poisson counting p-value $> 10^{-3}$) in both adjacent sidebands.
Starting with a two-bin window, the algorithm increases the signal window and shifts its location until all possible bin ranges, up to half the mass range spanned by the data, have been tested.  
The most significant departure from the smooth spectrum, defined by the set of bins that have the smallest probability of arising from a background fluctuation assuming Poisson statistics, is therefore identified.
The algorithm naturally accounts for the ``trials factor'' to assess the significance of its finding, by performing a series of pseudo-experiments to determine the probability that random fluctuations in the background-only hypothesis would create an excess as significant as the observed one anywhere in the spectrum.
The background to which the data are compared is obtained from the aforementioned fit, excluding the region with the biggest local excess of data in cases where the $\chi^2$\ test yields a p-value less than 0.01.  Although this is not the case in the actual data, it can happen in some of the pseudo-experiments that are used to determine the p-value.  
The reason for this exclusion is to prevent potential new physics signal from biasing the background.

The most significant discrepancy identified by the \BumpHunter\ algorithm is a three-bin excess in the dijet mass interval 995-1253 GeV.
The p-value of observing an excess at least as large as this assuming a background-only hypothesis is 0.39.  
We therefore conclude that there is no evidence for a resonance signal in the $\mjj$\ spectrum, and proceed to set limits on various models.

\subsection{Exclusion Limits Using the Dijet Mass}
We set Bayesian credibility intervals by defining a posterior probability density from the likelihood function for the observed mass spectrum, obtained by a fit to the background functional form and a signal shape derived from MC calculations.
A prior constant in the possible signal strength is assumed.
The posterior probability is then integrated to determine the 95\%\ credibility level (C.L.)  for a given range of models, usually parameterised by the mass of the resonance.  
A Bayesian approach is employed for setting limits using the dijet mass distribution as it simplifies the treatment of systematic uncertainties.

The systematic uncertainties affecting this analysis arise from instrumental effects, such as the jet energy scale (JES) and resolution (JER) uncertainties, the uncertainty on the integrated luminosity and the uncertainties arising from the background parameterization.
Extensive studies of the performance of the detector using both data and MC modelling have resulted in a JES uncertainty ranging from 3.2 to 5.7\%\ in the current data sample \cite{ATLAS:JES3}.
The systematic uncertainty on the integrated luminosity is 11\%\  \cite{ATLASintegLumi}.
The uncertainties on the background parameterization are taken from the fit results discussed earlier, and range from 3\%\ at 600~GeV to $\sim 40$\%\ at 3500~GeV.  
These uncertainties are incorporated into the analysis by varying all the sources according to Gaussian probability distributions and convolving these with the Bayesian posterior probability distribution. 
Credibility intervals are then calculated numerically from the resulting convolutions.

Uncertainties on the signal models come primarily from our choice of PDFs and 
the tune for the \Pythia\  MC, which provides the best match of observed 
data with the predictions with that choice of PDF.  
Our default choice of PDFs for the dijet mass analysis is MRST2007LO* \cite{Sherstnev:2007nd}\ with the MC09 tune \cite{ATLAS_MC09}.  
Limits are quoted also using CTEQ6L1 and CTEQ5L PDF sets, which provide an 
alternate PDF parametrization  and allow comparisons with previous 
results\cite{CDF:2009DijetSearch}, respectively.
For the $\qstar$\ limit analysis, we also vary the renormalization and factorization scales in the \Pythia\ calculation by factors of one-half and two, and find that the observed limit varies by $\sim0.1$~TeV. 

\subsection{Limits on Excited Quark Production}

The particular signal hypothesis used to set limits on excited quarks ($\qstar$)
has been implemented using the \Pythia\ MC generator, with fixed parameters to specify
the excited quark mass, $m_{\qstar}$\ and its decay modes, as discussed in Section~\ref{sec:theorymodels}.  Each choice of mass
constitutes a specific signal template, and a high-statistics set of MC events was created and
fully simulated for each choice of $m_{\qstar}$.  
The acceptance, $\Acc$, of our selection requirements ranges from 49\%\ to 58\%\ for $m_{\qstar}$\ from 600 to 3000~GeV, respectively.
The loss of acceptance comes mainly from the pseudorapidity requirements, which ensure that the candidate events have a high signal-to-background ratio.

\begin{figure}
  \centering {\includegraphics[width=0.60\textwidth]{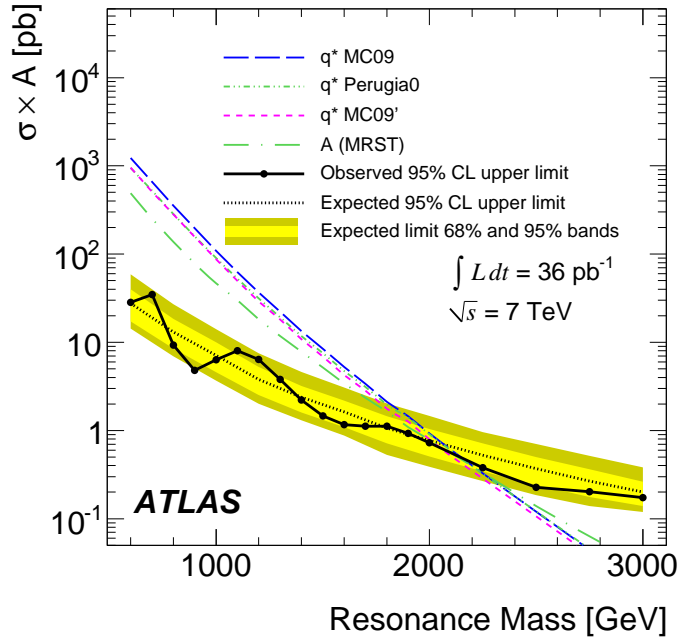}
            }
  \caption{The 95\%\ C.L. upper limits on the cross section times acceptance for a resonance decaying to dijets taking into account both statistical and systematic uncertainties (points and solid line) compared to an axigluon model and to a $\qstar$\ model with three alternate MC tunes.  We also show the expected limit (dotted line) and the 68\% and 95\% contours of the expected limit by the band.}
  \label{fig:qstarobsandexcl_wsyst}
\end{figure}

In Fig.~\ref{fig:qstarobsandexcl_wsyst}\ the resulting 95\%\ C.L.  limits on 
$\sigma\cdot {\cal A}$\ for excited quark production are shown as a function of the excited quark mass, 
where $\sigma$\ is the cross section for production of the resonance and $\Acc$\ is the 
acceptance for the dijet final state.
The expected limit is also shown, based on the statistics of the sample and assuming a background-only hypothesis.  
We see that the observed and expected limits are in reasonable agreement with each other, strengthening our earlier conclusion that there is no evidence of a signal above the smooth
background.
Comparing the observed limit with the predicted $\qstar$\ cross section times acceptance, we exclude at 95\% C.L. $\qstar$ masses in the interval $\MjjLowerExclusion < m_{\qstar} < \MjjLowerLimitMRST$~TeV.
The expected limit excludes $m_{\qstar} < \MjjLowerLimitMRSTEXPECTED$~TeV.

The sensitivity of the resulting limit to the choice of PDFs was modest, as shown in Table~\ref{Mjj exclusion limits}\ where the observed and expected mass limits are compared for several other models.  In all cases, the mass limits vary by less than 0.1~TeV.
The inclusion of systematic uncertainties result in modest reductions in the limit, illustrating that the limit setting is dominated by statistical uncertainties.

\begin{table*}
\caption{The 95\% C.L. lower limits on the allowed $\qstar$\ mass obtained
using different tunes and PDF sets.
The MC09$^\prime$ tune is identical to MC09
except for the \Pythia\ parameter PARP(82)$=2.1$ and use of the
CTEQ6L1 PDF set.}
\label{Mjj exclusion limits}
\begin{tabular}{ll|cc|cc}
~ & ~ & \multicolumn{2}{c}{Observed Limit [TeV]} & \multicolumn{2}{c}{Expected Limit [TeV]} \\
MC Tune & PDF Set  & Stat.~$\oplus$~Syst. & Stat.~only  & Stat.~$\oplus$~Syst. & Stat.~only \\ \hline
MC09~\cite{ATLAS_MC09} & MRST2007LO*~\cite{Sherstnev:2007nd} &   \MjjLowerLimitMRST & \MjjLowerLimitStatOnlyMRST & \MjjLowerLimitMRSTEXPECTED & \MjjLowerLimitStatOnlyMRSTEXPECTED \\
MC09$^\prime$
    &
  CTEQ6L1~\cite{Pumplin:2002vw} &
  \MjjLowerLimitCTEQsixLone & \MjjLowerLimitStatOnlyCTEQsixLone & \MjjLowerLimitCTEQsixLoneEXPECTED &
  \MjjLowerLimitStatOnlyCTEQsixLoneEXPECTED \\
Perugia0~\cite{Skands:2009zm} & CTEQ5L~\cite{Lai:1999wy} &
  \MjjLowerLimitCTEQfiveL & \MjjLowerLimitStatOnlyCTEQfiveL & \MjjLowerLimitCTEQfiveLEXPECTED &
  \MjjLowerLimitStatOnlyCTEQfiveLEXPECTED \\
\end{tabular}
\end{table*}

\subsection{Limits on Axigluon Production}

We set limits on axigluon production using the same procedure followed for the $\qstar$\ analysis, creating templates for the signal using the axigluon model described in Sec.~\ref{sec:theorymodels}\ and full detector simulation.
There are large non-resonant contributions to the cross section at low 
dijet mass, so we require at the parton-level that the axigluon invariant mass be between 0.7 and 1.3 times the nominal mass of the resonance.
Having made this requirement, we note that the axigluon and $\qstar$\ signal templates result in very similar limits.  So for convenience we use the $\qstar$\ templates in setting cross section limits on axigluon production.
 
The resulting limits are shown in Fig.~\ref{fig:qstarobsandexcl_wsyst}.
Using the MRST2007LO* PDFs, we exclude at 95\% C.L. axigluon 
masses in the interval $\MjjLowerExclusion < m < \MjjLowerLimitMRSTAxigluon$~TeV.
The expected limit is $m < \MjjLowerLimitMRSTEXPECTEDAxigluon$~TeV.
If only statistical uncertainties are included, the limit rises by $\sim~0.2$~TeV,
indicating that the systematic uncertainties are not dominant.

\subsection{Limits on Quantum Black Hole Production}

We search for production of Randall-Meade QBHs as these 
are expected to produce low multiplicity decays with a significant contribution to dijet final states.
Several scenarios are examined, with quantum gravity scales $M_D$\ ranging from 0.75~TeV to 4.0~TeV, 
and with the number of extra dimensions, $n$, ranging from two to seven.
The fully simulated MC events are used to create templates similar to the $\qstar$\ analysis.  
These QBH models produce threshold effects in $\mjj$\ with long tails to higher $\mjj$\ that  compete with the QCD background.  
However, the cross section is very large just above the threshold and so it is possible to extract 
limits given the resulting resonance-like signal shape.

\begin{figure}
  \centering \includegraphics[width=0.60\textwidth]{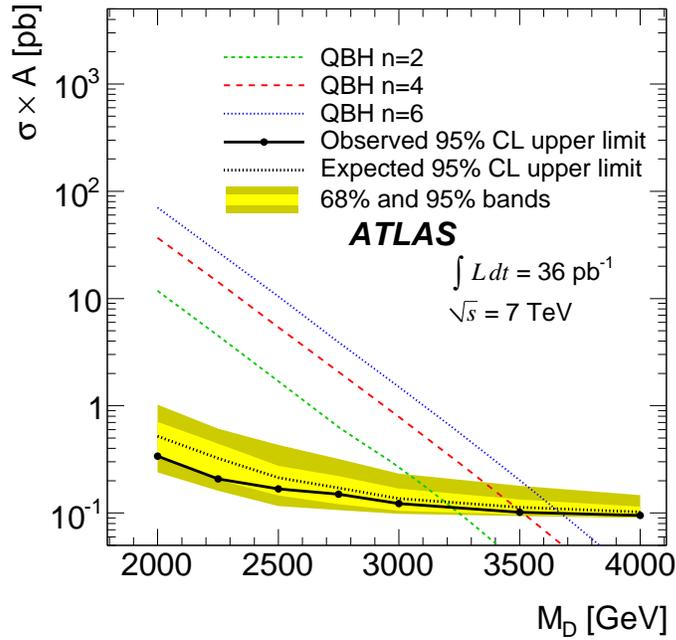}
  \caption{The 95\%\ C.L. upper limits on the cross section $\times$\ acceptance versus the quantum gravity
  mass scale $M_D$\  for a 
  Randall-Meade QBH model, taking into account both statistical and systematic uncertainties.
  The cross section $\times$\ acceptance for QBH models with two, four and six extra dimensions are shown.
  The 68\% and 95\% C.L. contours of the expected limit are shown as the band.}
  \label{fig:qbhobsandexcl_wsyst}
\end{figure}

The resulting limits are illustrated in 
Fig.~\ref{fig:qbhobsandexcl_wsyst}, showing the observed and expected limits, as well as the predictions for QBH production assuming two, four and six  extra dimensions.
The observed lower limits on the quantum gravity scale, $M_D$, with and without systematic uncertainty, and the expected limit
with and without systematic uncertainty, at 95\% C.L. are summarised in Table~\ref{tab:QBH}.
Using CTEQ6.6 parton distribution functions, we exclude at 95\% C.L. quantum gravity scales
in the interval $0.75 < M_D < \MjjLowerLimitMRSTQBHnsix$~TeV for the low multiplicity Randall-Meade QBHs with six extra dimensions.
The expected limit is $M_D < \MjjLowerLimitMRSTEXPECTEDQBHnsix$~TeV.

\begin{table*}
\caption[Mjj exclusion limits]{The 95\% C.L. lower limits on the allowed quantum gravity scale for various numbers
of extra dimensions.\label{tab:QBH}\\}
\begin{tabular}{c|cc|cc}
Number of  & \multicolumn{2}{c}{Observed $M_D$\ Limit [TeV]} & \multicolumn{2}{c}{Expected $M_D$\ Limit [TeV]} \\
Extra Dimensions & Stat.~$\oplus$~Syst. & Stat.~only  & Stat.~$\oplus$~Syst. & Stat.~only \\ \hline
2 & \MjjLowerLimitMRSTQBHntwo & \MjjLowerLimitStatOnlyMRSTQBHntwo & \MjjLowerLimitMRSTEXPECTEDQBHntwo & \MjjLowerLimitStatOnlyMRSTEXPECTEDQBHntwo \\
3 & \MjjLowerLimitMRSTQBHnthree & \MjjLowerLimitStatOnlyMRSTQBHnthree & \MjjLowerLimitMRSTEXPECTEDQBHnthree & \MjjLowerLimitStatOnlyMRSTEXPECTEDQBHnthree \\
4 & \MjjLowerLimitMRSTQBHnfour & \MjjLowerLimitStatOnlyMRSTQBHnfour & \MjjLowerLimitMRSTEXPECTEDQBHnfour & \MjjLowerLimitStatOnlyMRSTEXPECTEDQBHnfour \\
5 & \MjjLowerLimitMRSTQBHnfive & \MjjLowerLimitStatOnlyMRSTQBHnfive & \MjjLowerLimitMRSTEXPECTEDQBHnfive & \MjjLowerLimitStatOnlyMRSTEXPECTEDQBHnfive \\
6 & \MjjLowerLimitMRSTQBHnsix & \MjjLowerLimitStatOnlyMRSTQBHnsix & \MjjLowerLimitMRSTEXPECTEDQBHnsix & \MjjLowerLimitStatOnlyMRSTEXPECTEDQBHnsix \\
7 & \MjjLowerLimitMRSTQBHnseven & \MjjLowerLimitStatOnlyMRSTQBHnseven & \MjjLowerLimitMRSTEXPECTEDQBHnseven & \MjjLowerLimitStatOnlyMRSTEXPECTEDQBHnseven \\
\end{tabular}
\end{table*}

\subsection{Limits on RS Graviton Production}

We search for production of Randall-Sundrum gravitons by creating dijet mass templates using the MC calculation described in Sec.~\ref{sec:theorymodels}.
In this case, the sensitivity of the search is reduced by the lower production cross section, and by our kinematic criteria that strongly select for final states that have either high-energy hadronic jets or electromagnetic showers.

\begin{figure}
  \centering \includegraphics[width=0.60\textwidth]{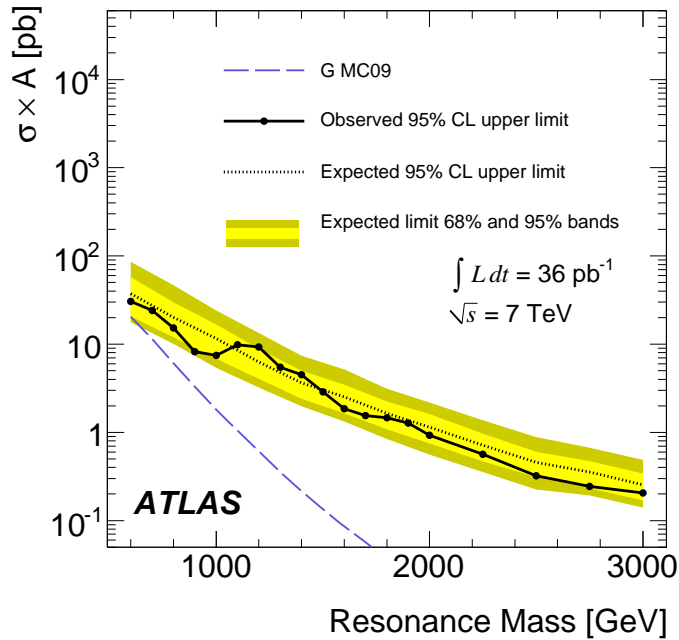}
  \caption{The 95\%\ C.L. upper limits on the cross section $\times$\ acceptance for a 
  Randall-Sundrum graviton, taking into account both statistical and systematic uncertainties.
  The 68\% and 95\% C.L. contours of the expected limit are shown as the band.}
  \label{fig:rsgobsandexcl_wsyst}
\end{figure}

The limits obtained for this hypothesis are illustrated in 
Fig.~\ref{fig:rsgobsandexcl_wsyst}, showing the observed and expected limits, as well as the predictions for RS graviton production.
It is not possible to exclude any RS graviton mass hypothesis, given the small expected signal 
rates and the relatively large QCD backgrounds.
A limit on RS graviton models could be established with increased statistics, though more sophisticated stategies to improve signal-to-background may be necessary.

\subsection{Simplified Gaussian Model Limits}

We have used these data to set limits in a more model-independent way by employing as our signal template a Gaussian profile with means ranging from  600 GeV to 4000~GeV and with the width, $\sigma$, varying from 3\% to 15\% of the mean.

Systematic uncertainties are treated in the same manner as described previously, using 
pseudo-experiments to marginalise the posterior probabilities that depend on
parameters that suffer from systematic uncertainty.
However, given that the decay of the dijet final state has not been modelled, assuming only that the resulting dijet width is Gaussian in shape, we adjusted the treatment of the jet energy scale by 
modelling it as an uncertainty in the central value of the Gaussian signal.

The 95\%\ C.L. limits are shown 
in Table~\ref{tab:GaussianLimits}, expressed 
in terms of number of events observed after all event selection criteria have been applied.
We stress that these event limits are determined by assuming a Gaussian signal shape.
Their variation as a function of mass and width reflects the statistical fluctuations of
data in the binned $\mjj$\ distribution used to set them.

These limits can be employed by computing for a given model the acceptance $\Acc$\ 
using a standard Monte Carlo calculation.
The jet \pT\ and $\eta$\ requirements should first be applied to determine the expected signal
shape in $m_{jj}$.  
Since a Gaussian signal shape has been assumed in determining the limits, we recommend 
removing any long tails in $\mjj$\ (a $\pm20$\%\ mass window is recommended). 
The fraction of MC events surviving these requirements  is an estimate of the acceptance, and can be used to calculate the expected event yield given a cross section for the process and assuming a sample size of \integLumi.
This event yield can then be compared with the limit in Table~\ref{tab:GaussianLimits}, matching the expected signal mean and width to the appropriate entry in the table.

\begin{table}
\begin{center}
\caption[]{The 95\% C.L. upper limits on the number of observed signal events for Gaussian reconstructed $\mjj$\
distributions.  The effects of systematic uncertainties due to the luminosity, the background fit and the jet energy scale have been included.  We present the signal widths as $\sigma/m$.}
\label{tab:GaussianLimits}
\begin{tabular}{cccccc}
\hline
~ & \multicolumn{5}{c}{$\sigma/m$} \\
Mean $m$ (GeV) & 0.03 & 0.05 & 0.07 & 0.10 & 0.15 \\ \hline
600   &  434  &  638  &  849  &  1300  &  1990  \\
700   &  409  &  530  &  789  &  1092  &  945  \\
800   &  173  &  194  &  198  &  218  &  231  \\
900   &  88  &  103  &  123  &  162  &  311  \\
1000   &  147  &  179  &  210  &  278  &  391  \\
1100   &  143  &  169  &  204  &  263  &  342  \\
1200   &  91  &  120  &  168  &  223  &  262  \\
1300   &  65  &  80  &  101  &  120  &  122  \\
1400   &  35  &  42  &  50  &  57  &  66  \\
1500   &  24  &  27  &  32  &  40  &  60  \\
1600   &  21  &  25  &  29  &  36  &  49  \\
1700   &  26  &  27  &  28  &  38  &  43  \\
1800   &  25  &  26  &  30  &  32  &  34  \\
1900   &  22  &  22  &  25  &  25  &  26  \\
2000   &  13  &  16  &  19  &  19  &  17  \\
2100   &  10  &  12  &  14  &  16  &  17  \\
2200   &  8.4  &  9.4  &  11  &  10  &  11  \\
2300   &  6.8  &  7.3  &  7.4  &  8.3  &  9.0  \\
2400   &  4.9  &  5.2  &  6.1  &  6.6  &  8.0  \\
2500   &  4.6  &  4.9  &  5.4  &  6.4  &  6.9  \\
2600   &  4.9  &  5.0  &  5.3  &  6.0  &  6.6  \\
2700   &  5.1  &  5.0  &  5.0  &  5.2  &  5.7  \\
2800   &  5.0  &  5.0  &  4.9  &  5.0  &  5.2  \\
2900   &  4.6  &  4.5  &  4.7  &  4.6  &  4.8  \\
3000   &  4.1  &  4.2  &  4.3  &  4.5  &  4.7  \\
3200   &  3.2  &  3.5  &  3.6  &  3.8  &  4.1  \\
3400   &  3.1  &  3.1  &  3.2  &  3.5  &  3.7  \\
3600   &  3.1  &  3.1  &  3.1  &  3.3  &  3.6  \\
3800   &  3.1  &  3.1  &  3.1  &  3.2  &  3.3  \\
4000   &  3.1  &  3.1  &  3.1  &  3.1  &  3.3  \\ 
\hline
\end{tabular}
\end{center}
\end{table}

% ------------------------------------  Dijet Angular Distribution Search ----------------------------------------%

\section{Angular Distribution Analyses}
\label{sec:Ang}

For all angular distributions analyses, the common event selection criteria described
in Sec.~\ref{sec:selection}\ are applied, including the transverse momentum requirements
on the two leading jets: $\mpT^{j_1} > 60$ GeV and $\mpT^{j_2} > 30$ GeV.
Additionally, $\chi$ distributions are accumulated only for events that satisfy
$|y_B|  < 1.10$\ and $|y^*| < 1.70$.
The $|y^*|$ criterion determines the maximum $\chi$\ of 30 for this analysis.
These two criteria limit the rapidity range of both jets to $|y_{1,2}| < 2.8$\ and 
define a region within the space of accessible $y_1$ and $y_2$\ with full and uniform 
acceptance in $\chi$\ for $\mjj > 500$~GeV.
These kinematic cuts have been optimised by MC studies of QCD and new physics
signal samples to assure high acceptance for all dijet masses.

Detector resolution effects smear
the $\chi$ distributions, causing events to migrate between neighboring bins.
This effect is reduced by choosing the $\chi$ bins to match the natural 
segmentation of the calorimeter, making them intervals of constant $\Delta y$\ for these high \pT\ dijet events. 
The $\mFchi$\ and \Fchimjj\ variables are even less sensitive to migration effects, 
given that they depend on separation of the data sample into only two $\chi$\ intervals.

\subsection{Systematic and Statistical Uncertainties}

Dijet angular distribution analyses have a reduced sensitivity to the JES and JER
uncertainties compared to other dijet measurements since data and theoretical 
distributions are normalised to unit area for each mass bin in all cases.  
Nevertheless, the JES still represents the dominant systematic uncertainty
in the current studies.

As described in a previous publication~\cite{ATLAS:2010eza}, our dijet angular analyses use
pseudo-experiments to convolve statistical, systematic and theoretical uncertainties.
The primary sources of theoretical uncertainty are NLO QCD renormalization ($\mu_R$) and factorization scales ($\mu_F$), and PDF uncertainties.
The former are varied by a factor of two independently,
while the PDF errors are sampled from a Gaussian distribution determined using 
CTEQ6.6 (NLO) PDF error sets.  The resulting bin-wise uncertainties for normalised $\chi$ distributions are typically up to 3\% for the combined NLO QCD scales and 1\% for
the PDF uncertainties.  These convolved experimental and theoretical uncertainties are calculated
for all Monte Carlo angular distributions (both QCD and new physics samples).
These statistical ensembles are used for estimating p-values when comparing QCD predictions
to data, and for parameter determination when setting limits.

\subsection{Observed $\chi$\ and \Fchimjj\ Distributions}

\begin{figure}[t]
  \centering \includegraphics[width=0.60\textwidth]{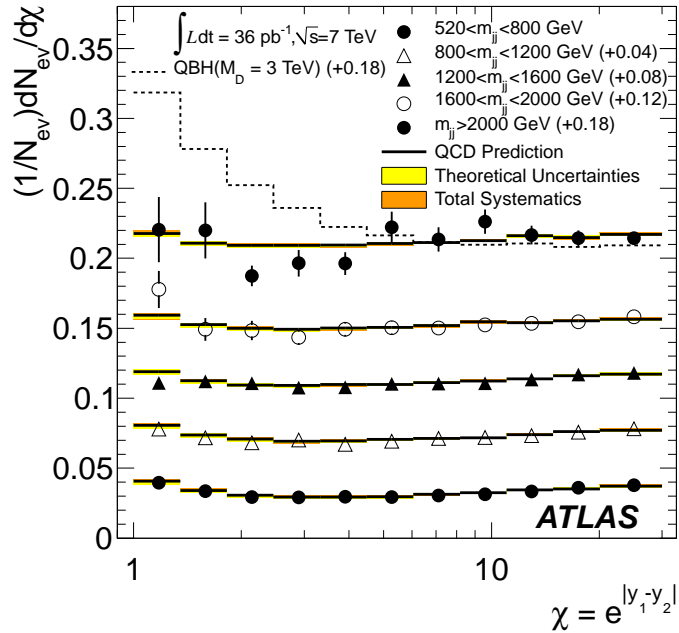}
  \caption{The $\chi$ distributions for $520 < \mjj < 800$ GeV, $800 < \mjj < 1200$ GeV,
      $1200 < \mjj< 1600$ GeV, $1600 < \mjj < 2000$ GeV, and $\mjj > 2000$ GeV. 
      Shown are the QCD predictions with systematic uncertainties (narrow bands),
      and data points with statistical uncertainties. The dashed line is the prediction for
      a QBH signal for $M_D =3$~TeV and $n = 6$\ in the highest mass bin.
      The distributions and QCD predictions have been offset by the amount shown in the 
      legend to aid in 
      visually comparing the shapes in each mass bin.}
  \label{fig:chisvsQCD}
\end{figure}

The analysis method used in the first ATLAS publication on this topic~\cite{ATLAS:2010eza} is 
revisited here for the full 2010 data sample. 
The $\chi$\ distributions are shown in Fig.~\ref{fig:chisvsQCD}\ for several
relatively large $\mjj$\ bins, defined by the bin boundaries of 520, 800, 1200, 1600 and 2000~GeV.
There are 71,402 events in the sample, ranging from 42,116 events in the lowest mass bin to  212 events with $\mjj > 2000$~GeV.  These bins were chosen to assure sufficient statistics in each mass bin.  This is
most critical for the highest mass bin - the focal point for new physics searches.
The $\chi$\ distributions are compared in the figure to the predictions from QCD MC models and the signal that would be seen in one particular new physics model, a QBH scenario with a quantum gravity mass scale of 3~TeV and six extra dimensions.  

The data appear to be consistent with the QCD predictions, which include systematic uncertainties.
To verify this, a binned likelihood is calculated for each distribution assuming that the sample
consists only of QCD dijet production.
The  expected distribution of this
likelihood is then calculated using pseudo-experiments drawn from the QCD MC sample and convolved with the systematic uncertainties as discussed above.
The p-values for the observed likelihood values, from the lowest to
highest mass bins, are 0.44, 0.33, 0.64, 0.89 and 0.44, respectively, confirming that the SM
QCD hypothesis is consistent with the data.

\begin{figure}[t]
  \centering \includegraphics[width=0.60\textwidth]{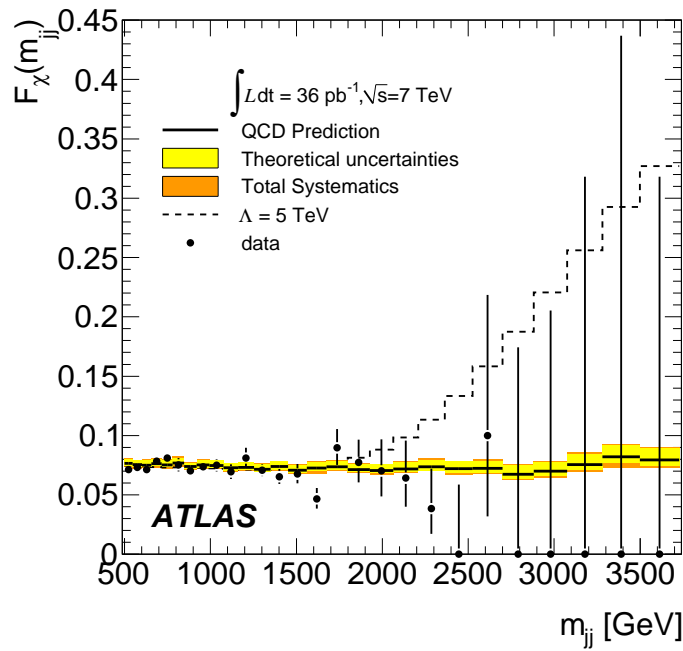}
  \caption{The \Fchimjj\ function versus $\mjj$.  We show  
 the QCD prediction with systematic uncertainties (band),  and data points (black points) with statistical uncertainties. 
The expected signal from QCD plus a quark contact interaction with
      $\Lambda$ = 5.0 TeV is also shown.}
  \label{fig:fchicompQCD}
\end{figure}

We compute the \Fchimjj\ observable, introduced in Sec.~\ref{sec:kinematics}, using the same
mass binning employed in the dijet resonance searches.
The observed \Fchimjj\ data are shown in Fig.~\ref{fig:fchicompQCD}\ and 
compared to the QCD predictions, which include systematic uncertainties.  
We also show the expected behaviour of \Fchimjj\ if a contact interaction
with the compositeness scale $\Lambda=5.0$~TeV were present. 
Statistical analyses using \Fchimjj\ use mass bins starting at 1253~GeV to be 
most sensitive to the high dijet mass region.
Assuming only QCD processes and including systematic uncertainties,
the p-value for the observed binned likelihood is 0.28, indicating that
these data are consistent with QCD predictions.

In the absence of any evidence for signals associated with new physics phenomena, these distributions
are used to set 95\%\ confidence level (C.L.) exclusion limits on a number of new physics hypotheses.

\subsection {Exclusion Limits from Likelihood Ratios}

Most of the dijet angular distribution analyses described below use
likelihood ratios for comparing different hypotheses and parameter estimation.
Confidence level limits are set using the frequentist CL$_{\mathrm{s+b}}$ approach \cite{Junk:1999}.  
As an example, for the \Fchimjj\ distributions the variable $Q$ is defined
as follows:
\begin{eqnarray}
Q = - 2\left[\ln L\left(\mFchimjj | H0\right) - \ln L\left(\mFchimjj | H1\right)\right],
\label{eg:loglikelihoodratio}
\end{eqnarray}
where $H0$ is the null hypothesis (QCD only), $H1$ is a specific hypothesis
for new physics with fixed parameters and $L(\mFchimjj | H)$\ is the binned 
likelihood for the \Fchimjj\ distribution assuming $H$\ as the hypothesis. 
Pseudo-experiments are used to 
determine the expected distribution for $Q$\ for specific hypotheses.
The new physics hypothesis is then varied to calculate a Neyman
confidence level.

\subsection{Limits on Quark Contact Interactions}

The \Fchimjj\ variable is used for the first time in this paper
to set limits on quark contact interactions (CI), as described in
Section~\ref{sec:theorymodels}.
MC samples of QCD production modified by a contact interaction are
created for values of $\Lambda$\ ranging from 0.50 to 8.0~TeV.  

For the pure QCD sample (corresponding to $\Lambda = \infty$), the \Fchimjj\
distribution is fit to a 2nd order polynomial. 
For MC samples with finite $\Lambda$, the distributions are fit, as a
function of $\mjj$, to the 2nd order polynomial plus a Fermi function,
which is a good representation of the onset curve for contact interactions.
QCD K-factors from Section~\ref{sec:theorymodels}\ are applied to the QCD-only component
of the spectra before calculating \Fchimjj.
This is done through an approximation that neglects possible NLO
corrections in the interference term between the QCD matrix element
and the contact interaction term.  
The issue of NLO corrections to contact terms has been independently
identified elsewhere~\cite{Gao:2011}.

\begin{figure}[t]
  \centering \includegraphics[width=0.60\textwidth]{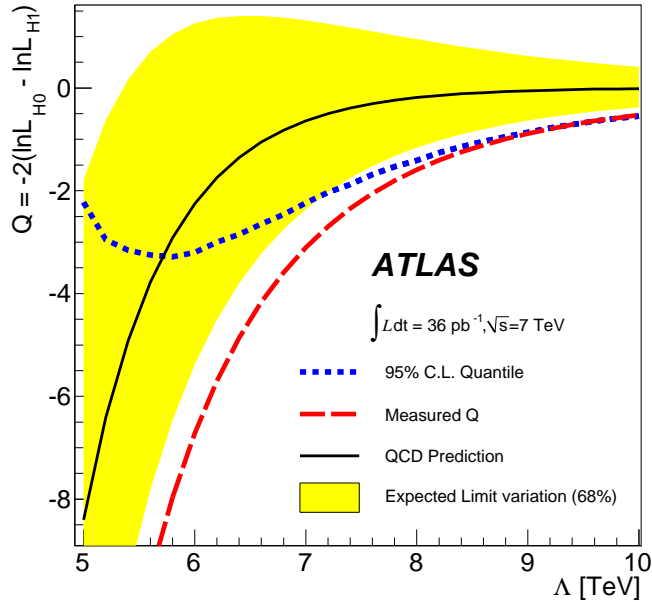}
  \caption{The log-likelihood ratio defined by the \Fchimjj\ distribution versus the strength of the contact interaction, $\Lambda$.
  The contact interaction limit is set comparing the measured log-likelihood ratio with that expected for a given value of $\Lambda$.}
  \label{fig:CIlimfchi}
\end{figure}

The \Fchimjj\ event sample is fit in each $\mjj$\ bin of the
distribution as a function of $1/\Lambda^2$, creating a predicted
\Fchimjj\ surface as a function of $\mjj$\ and $\Lambda$.  This surface
enables integration in $\mjj$\ vs $\Lambda$ for continuous values of $\Lambda$. 
Using this surface, the 95\% C.L. limit on $\Lambda$\ is determined
using the log-likelihood ratio defined in Eq.~\ref{eg:loglikelihoodratio}.
The resulting 95\% C.L. quantile is shown in Fig.~\ref{fig:CIlimfchi}. 

Figure~\ref{fig:CIlimfchi} also shows the expected value of $Q$\ for various
choices of $\Lambda$\ as well as the expected 95\%\ C.L. limit and its 68\%\
contour interval.

The observed exclusion limit is found from the point where
the 95\% quantile (dotted line) crosses the median value of
the distribution of Q values for the QCD prediction (dashed
line). This occurs at $\Lambda = \FchimjjLambda$~TeV.
The expected limit is $\Lambda = \FchimjjEXPECTEDLambda$~TeV.
The observed result is significantly
above the expected limit because the data has fewer centrally produced, high mass dijet events than expected from QCD alone, as can be seen in 
Fig.~\ref{fig:fchicompQCD}\ where the observed values of \Fchimjj\ fall below the QCD
prediction for dijet masses around 1.6 TeV and above 2.2~TeV.
These data are statistically compatible with QCD, as evidenced by the p-value of the binned 
likelihood.  
The expected probability that a limit at least as strong as this would be observed is $\sim 8$\%.

As a cross-check, a Bayesian analysis of \Fchimjj\ has been
performed, assuming a prior which is constant in 1/$\Lambda^2$.
This analysis sets a 95\%\ credibility level of $\Lambda > 6.7$~TeV.
The expected limit from this Bayesian analysis is 5.7~TeV,
comparable to the CL$_{\mathrm{s+b}}$ expected limit.  While the
observed limit from CL$_{\mathrm{s+b}}$ analysis is significantly higher
than the Bayesian results, we
have no basis on which to exclude the CL$_{\mathrm{s+b}}$ result a posteriori.

As an additional cross-check, the earlier $\mFchi$\ analysis of the $dN/d\chi$\ distributions,
coarsely binned in $\mjj$~\cite{ATLAS:2010eza}, has been repeated.
With the larger data sample and higher threshold on the highest $\mjj$\ bin (2 TeV), the 
observed and expected limits are $\Lambda > \FchiKFacQCDCILambda$~TeV and $\Lambda > \FchiKFacQCDCIEXPECTEDLambda$~TeV, respectively.
As anticipated, these limits are not as strong as those arising from the \Fchimjj\
analysis because of the coarser $\mjj$\ binning.

Finally, an analysis was performed to see if a more sensitive measure
could be created by setting limits based on all 11 bins of the highest
mass ($m_{jj} > 2$ TeV) $\chi$ distribution, instead of the two intervals used in the $\mFchi$\
analysis.  In this method, for each bin the same interpolating function used in
the \Fchimjj\ analysis is fit to the bin contents resulting from all QCD+CI MC samples,
yielding the CI onset curve.  Limits are set  using the same log-likelihood ratio and
pseudo-experiment methods employed in the \Fchimjj\ analysis.  The observed 95\%\ C.L.
limit is $\Lambda > \ElevenBinChiLambda$~TeV.   For the current data sample,
the expected limit is \ElevenBinChiEXPECTEDLambda~TeV.
Since the expected limit exceeds that from the $\mFchi$\ analysis, this method
shows promise for future analyses.

\subsection{Limits on Excited Quark Production}

\begin{figure}[t]
  \centering \includegraphics[width=0.60\textwidth]{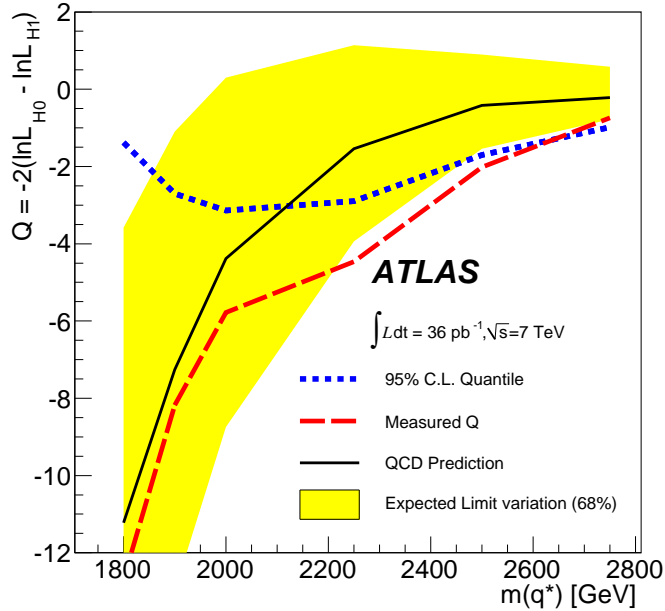}
  \caption{The 95\%\ C.L. limits on the excited quark model using the 
  logarithm of the likelihood ratios obtained from the \Fchimjj\ distribution.
  The expected 68\%\ interval for the expected limits are shown by the band.}
  \label{fig:qstarlimfchi}
\end{figure}

The \Fchimjj\ distributions are also used to set limits
on excited quark production. As described earlier, the $\qstar$ 
model depends only on the single parameter, $m_\qstar$.
Twelve simulated $\qstar$ mass ($m_{\qstar}$) samples in the range from
1.5 to 5.0 TeV are used for the analysis.
Based on the assumption that interference of QCD with excited quark resonances is negligible, 
$\qstar$ MC samples are scaled by their cross sections and
added to the NLO QCD sample (which has been corrected using bin-wise K-factors).
By analogy with the contact interactions analysis above, a likelihood is constructed
by comparing the expected and observed \Fchimjj\ distributions for each value of
$m_{\qstar}$.  We then form a likelihood ratio with respect to the QCD-only
hypothesis and use this to set confidence intervals on the production of a $\qstar$.

Figure~\ref{fig:qstarlimfchi} illustrates the limit setting procedure for the 
$\qstar$~model. The observed exclusion limit is found from the point where
the 95\% quantile (dotted line) crosses the measured value of Q (dashed line).
This occurs for $m_{\qstar} = \FchimjjQstar$~TeV.  
The expected limit, determined from the point where the QCD prediction
(solid line) crosses the 95\% quantile, is \FchimjjEXPECTEDQstar~TeV.  
The observed limit falls near the 68\% ($\pm1\  \sigma$) interval of
the expected limit.
The difference in observed and expected limits arises from the lower observed \Fchimjj\ values for dijet masses above 2.2~TeV.

This result can be compared to the limits obtained from the dijet resonance analysis, 
which sets observed exclusion limits on $\qstar$\ 
masses of \MjjLowerLimitMRST~TeV.

\subsection{Limits on $\sigma_{QCD} \times \Acc_{QCD}$ for Additive Signals}

For new physics signals that do not interfere significantly with QCD,
limit setting may be done in a more model-independent way.
Monte Carlo signal samples are simulated independently from QCD samples and,
for any given choice of new physics model parameters, the two samples are
added to create a combined MC sample for comparison with data.
This is implemented by introducing a variable $\thetanp$ defined as
\begin{eqnarray}
\thetanp = \frac{\sigma_{np} \times \Acc_{np}}{\sigma_{QCD} \times \Acc_{QCD}},
\end{eqnarray}
where $\sigma$\ and $\Acc$\ are the cross section and acceptance for the given 
process, and ``$np$'' refers to the new physics process.
This variable represents the contribution of signal events in terms of cross section times
acceptance relative to the QCD background.  
The acceptance factors are determined by MC calculations.

\subsection{Limits on Quantum Black Hole Production}

The Randall-Meade model of QBH production~\cite{RandallMeade},  
introduced in Section~\ref{sec:theorymodels}\ and used to set limits
on these phenomena in the dijet resonance analysis,  
is employed again here to search for QBH production using the dijet angular 
distributions.
The $\thetanp$-parameter limit-setting method, sensitive to
$\sigma_{QBH} \times \Acc_{QBH}$, is used for this analysis since the QBH
production model does not include interference with QCD.

MC samples are created corresponding to discrete values of the QBH
quantum gravity mass scale $M_D$\ ranging from 2.0 to 4.0~TeV and
for two to seven extra dimensions ($n$), and are used to determine the
acceptance $\Acc_{QBH}$. The acceptance is found to  
vary from 58\%\ to 89\%\ as $M_D$\ is varied from 2.0 to 4.0~TeV,
for the case of six extra dimensions.  These studies have shown that
the signal acceptance for the model considered here varies
little with the model parameter $n$.  Thus, $A_{QBH}$ found from full
simulation of the sample with $n=6$\ is applied to the limit analysis for other 
choices of $n$, which have different cross sections.

The MC events with dijet masses greater than 2.0 TeV
are binned in $\chi$ with the same bin boundaries used in Fig.~\ref{fig:chisvsQCD}.
Pseudo-experiments are used to incorporate the JES uncertainty into the predicted $\chi$\ 
distributions. In each $\chi$ bin a linear fit is made for $dN/d\chi$ vs $\thetanp$, creating
a family of lines that define a $dN/d\chi$ surface in $\thetanp$ vs $\chi$.
Scale and PDF uncertainties, and the uncertainty on the JES correlation between the two jets, 
are incorporated into this surface using pseudo-experiments, and a value of \Fchi\ is
calculated from each distribution.  The expected distribution of \Fchi\ values is obtained using
additional pseudo-experiments modelling the finite statistics of the high $\mjj$\ event sample.
A likelihood ratio is formed comparing the likelihood of a given value of \Fchi\
for a QBH hypothesis to the likelihood from QCD processes alone.  
Finally, pseudo-experiments are performed to extract the 95\%\ C.L. exclusion limit
on \Fchi\ as a function of $\thetanp$.

\begin{figure}[t]
  \centering \includegraphics[width=0.60\textwidth]{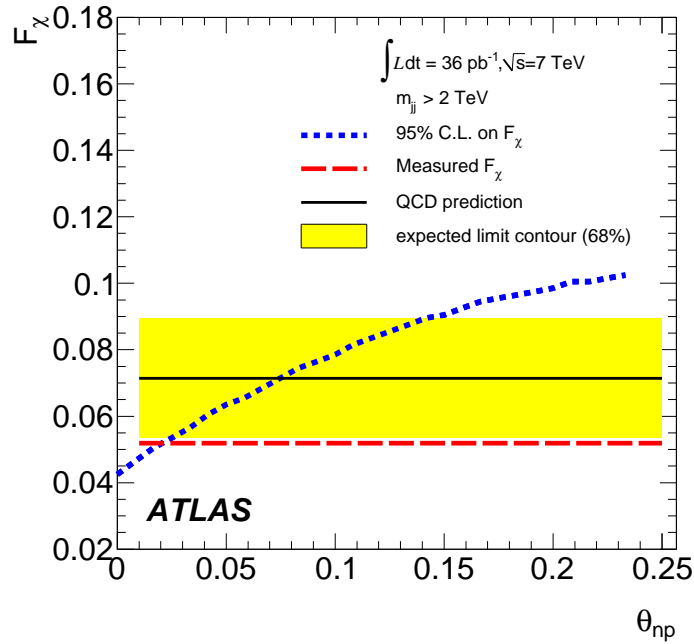}
  \caption{The 95\%\ C.L. upper limits in the $\mFchi-\thetanp$\ plane for  QBH production
  for $n=6$\ extra dimensions.}
  \label{fig:QBHlimits}
\end{figure}

Figure~\ref{fig:QBHlimits} illustrates the $\thetanp$ parameter limit-setting procedure
for the case $n = 6$. The observed exclusion limit is found
from the point where the 95\% C.L. contour (dotted line) crosses the measured value of 
$\mFchi = 0.052$\ (dashed line), which occurs at $\thetanp$ = 0.020.  
The expected limit, determined from the point where
the QCD prediction, 0.071 (solid line), crosses the 95\% C.L. contour, is at 0.075.  
The observed limit falls just outside the 68\% ($\pm 1 \sigma$) interval of  the expected limit.
These limits on $\thetanp$\ are translated into limits on $\sigma \times \Acc$ using
the QCD cross section and the acceptance for fully simulated dijets,
$\sigma_{QCD} \times \Acc_{QCD} = 7.21$~pb, resulting in an observed 95\%\ C.L. upper limit 
$\sigma_{QBH} \times \Acc_{QBH} < 0.15$~pb.

\begin{figure}[t]
  \centering \includegraphics[width=0.60\textwidth]{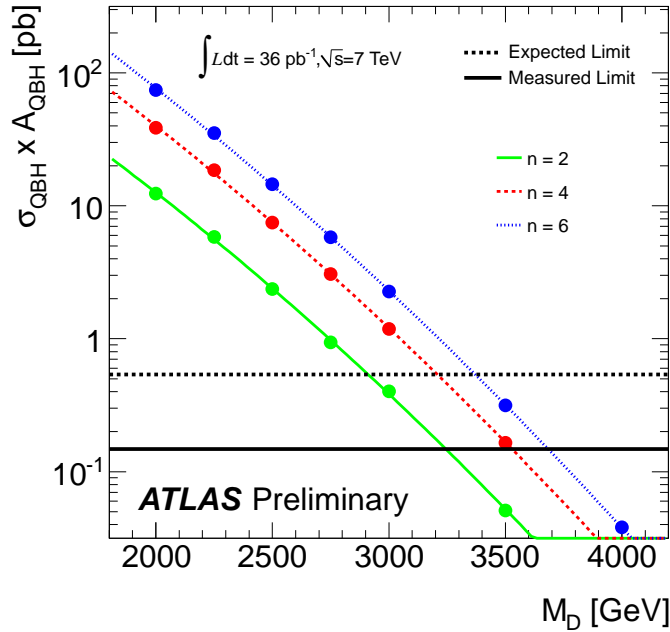}
  \caption{The cross section $\times$ acceptance for QBHs as a function of $M_D$\ for two, four and six extra dimensions.
  The measured and expected limits are shown in the solid and dashed line.}
  \label{fig:limsetMD}
\end{figure}

Figure~\ref{fig:limsetMD} shows the $\sigma_{QBH} \times \Acc_{QBH}$\ vs $M_D$\ curves for 
two, four and six extra dimensions.
The measured and expected limits for $\sigma_{QBH} \times \Acc_{QBH}$ are plotted
as horizontal lines.  The crossing points of these lines with the $n$\ vs $M_D$\ curve  yield
expected and observed exclusion limits for the QBH model studied here.
The 95\%\ C.L. lower limit on the quantum gravity mass scale is  \FchiLowerLimitMRSTQBHnsix~TeV for six extra dimensions.   The expected limit is 
\FchiLowerLimitMRSTEXPECTEDQBHnsix~TeV.
The limits for all extra dimensions studied here, $n=2$ to $7$, are listed in Table~\ref{tab:QBHlimstabl}.

\begin{table}[t]
\caption{The 95\%\ C.L. exclusion limits on $M_D$\ for various choices of extra dimensions for
the Randall-Meade QBH model determined by the $\thetanp$ parameter analysis for $\mjj > 2.0$~TeV.} 
\label{tab:QBHlimstabl}
\centering
\begin{tabular}{ccc}
	\hline
 	 $n$ & Expected  & Observed \\
	 Extra Dimensions  &  Limit (TeV)  &  Limit (TeV) \\
	\hline 
		2 &	\FchiLowerLimitMRSTEXPECTEDQBHntwo	& \FchiLowerLimitMRSTQBHntwo  \\
		3 &	\FchiLowerLimitMRSTEXPECTEDQBHnthree	& \FchiLowerLimitMRSTQBHnthree  \\
		4 &	\FchiLowerLimitMRSTEXPECTEDQBHnfour	& \FchiLowerLimitMRSTQBHnfour  \\
		5 &	\FchiLowerLimitMRSTEXPECTEDQBHnfive	& \FchiLowerLimitMRSTQBHnfive  \\
		6 &	\FchiLowerLimitMRSTEXPECTEDQBHnsix	& \FchiLowerLimitMRSTQBHnsix  \\
		7 &	\FchiLowerLimitMRSTEXPECTEDQBHnseven	& \FchiLowerLimitMRSTQBHnseven  \\
	\hline
\end{tabular}
\end{table}

The limit for $\sigma_{QBH} \times \Acc_{QBH}$ may also be applied to any
new physics model that satisfies the following criteria: (1) the \Fchi\ of
$np$ signal-only event samples should be roughly independent of $m_{jj}$, as is the
case for $\qstar$, QBH, and contact interactions; and (2) this \Fchi\ should
be close to the value of \Fchi\ for the current QBH study [0.58]. It is not necessary
that the $m_{jj}$ spectrum be similar, or that the QCD+$np$ sample have the
same \Fchi. 

It should also be noted that the results from this $\thetanp$ parameter analysis
are in agreement with the expected and observed limits obtained for the same
QBH model in the dijet resonance analysis. These two analyses are focusing on
complementary variables in the two-dimensional space of $\mjj$\ and $\chi$\ yet arrive at
similar limits.

A cross-check of these results is made by extracting a QBH limit using the \Fchimjj\ distribution
for the case of six extra dimensions.
Signal and background samples are created by combining the 
QBH signals for various $M_D$'s with the QCD background sample corrected by K-factors.
The \Fchimjj\ distribution is then fit in each $\mjj$\ bin as a function of $1/{M_D}^2$\
using the same interpolating function employed in \Fchimjj\ contact interactions 
analysis. The likelihood ratio construction and limit setting procedures used 
in the CI analysis are also applied in this study, resulting in observed and 
expected 95\%\ C.L. limits for $M_D$\ of \FchimjjQBH~TeV and \FchimjjEXPECTEDQBH~TeV,
respectively.
A further cross-check is performed using the 11-bin $\chi$ analysis to set 
limits on a QBH for the case of six extra dimensions.
This study yields an observed 95\%\ C.L. limit of $M_D > \ElevenBinChiQBH$~TeV and
an expected limit of $M_D > \ElevenBinChiEXPECTEDQBH$~TeV. 

The expected and observed limits resulting from these four studies are summarised with the 
results of the other analyses in  Table~\ref{tab:Summary}.  
The strongest expected limits on QBH production come from the dijet resonance analysis, but the angular analyses are in close agreement, yielding limits within 0.3~TeV of each other for
the QBH hypothesis under study.

% ------------------------------------ Conclusions  ----------------------------------------%

\section{Conclusion}
\label{sec:conclusions}

Dijet mass and angular distributions have been measured by the ATLAS
experiment over a large angular range and spanning dijet masses
up to $\approx \HighestDijetMass$~TeV
using \integLumi\ of 7 TeV $pp$ collision data.
The angular distributions are in good agreement
with QCD predictions and we find no evidence for new phenomena.
Our analysis,  employing both the dijet mass and the dijet angular distributions, places
the most stringent limits on contact interactions, resonances and threshold phenomena to date.

\begin{table}[t]
\centering {
\caption {The 95\% C.L. lower limits on the masses and energy scales of the models 
examined in this study.  
We have included systematic uncertainties into the upper limits using the techniques described in the text.
The result with the highest expected limit is shown in bold face and is our quoted result.
}
\label{tab:Summary}
\begin{tabular}{lcc}
\hline
Model and Analysis Strategy  \quad \quad &  \multicolumn{2}{c}{95\%\ C.L. Limits (TeV)} \\
            &  Expected  &  Observed\\
      \hline
\multicolumn{3}{c}{Excited Quark $\qstar$ } \\
\hline
Resonance in $m_{jj}$   &   
\MjjLowerLimitMRSTEXPECTED & \MjjLowerLimitMRST     \\ 
{\bf \Fchimjj\ }   &  
{\bf \FchimjjEXPECTEDQstar}  & {\bf \FchimjjQstar} \\    
\hline \multicolumn{3}{c}{
Randall-Meade Quantum Black Hole for $n=6$} \\ \hline
{\bf Resonance in $m_{jj}$ } &  {\bf \MjjLowerLimitMRSTEXPECTEDQBH} &  {\bf \MjjLowerLimitMRSTQBH }   \\
\Fchimjj\              &   \FchimjjEXPECTEDQBH  &  \FchimjjQBH  \\    
$\thetanp$ Parameter for $\mjj > 2$~TeV  & 
   \FchiLowerLimitMRSTEXPECTEDQBHnsix & \FchiLowerLimitMRSTQBHnsix \\
11-bin $\chi$\ Distribution for $\mjj > 2$~TeV &
\ElevenBinChiEXPECTEDQBH & \ElevenBinChiQBH \\
\hline \multicolumn{3}{c}{
Axigluon
} \\ \hline
{\bf Resonance in $\mjj$ }  & 
{\bf \MjjLowerLimitMRSTEXPECTEDAxigluon} & {\bf \MjjLowerLimitMRSTAxigluon} \\  
\hline \multicolumn{3}{c}{
Contact Interaction $\Lambda$
} \\ \hline
{\bf \Fchimjj }  &   
{\bf \FchimjjEXPECTEDLambda}  & {\bf  \FchimjjLambda } \\
\Fchi\ for $\mjj > 2$~TeV  &  
\FchiKFacQCDCIEXPECTEDLambda  &  \FchiKFacQCDCILambda  \\ 
11-bin $\chi$\ Distribution for $\mjj > 2$~TeV &
\ElevenBinChiEXPECTEDLambda & \ElevenBinChiLambda \\
\hline
\end{tabular}
     } 
\end{table}

In Table~\ref{tab:Summary}\ the constraints on specific 
models of new physics that would contribute to dijet final states
are summarised.

We quote as the primary results the limits using 
the technique with the most stringent expected limit.
Therefore, we exclude at 95\% C.L. excited quarks with 
masses in the interval $\MjjLowerExclusion < m_{q^\ast} <  \FchimjjQstar$~TeV,
axigluons with masses between 0.60~TeV and \MjjLowerLimitMRSTAxigluon~TeV, and
Randall-Meade quantum black holes with  $0.75 < M_D <   \MjjLowerLimitMRSTQBH$~TeV assuming six extra dimensions.

We also exclude at 95\%\ C.L. quark contact interactions with a scale $\Lambda < \FchimjjLambda$~TeV. As noted earlier, the observed limit is significantly above the expected limit of \FchimjjEXPECTEDLambda~TeV for this data sample, and above the
limits from an alternative calculation using Bayesian statistics.  However, we quote this result since the statistical approach is a standard procedure that was chosen a priori.

In a number of cases, searches for the same phenomenon have been performed
using dijet mass distributions, dijet angular distributions, or both.
We are able to set comparable limits using these complementary techniques, while at the same time searching for evidence of narrow resonances, threshold effects, and enhancements in angular distributions that depend on the dijet invariant mass.  

This combined analysis is a sensitive probe into new physics that is expected to emerge at the TeV scale.  With increased integrated luminosity and continued improvements to analysis techniques and models, we expect to increase the ATLAS discovery reach for new phenomena that affect dijet final states.

% ------------------------------------ Acknowledgements  ----------------------------------------%

\section*{Acknowledgements}

We thank CERN for the very successful operation of the LHC, as well as the support staff from our institutions without whom ATLAS could not be operated efficiently.

We acknowledge the support of ANPCyT, Argentina; YerPhI, Armenia; ARC, Australia; BMWF, Austria; ANAS, Azerbaijan; SSTC, Belarus; CNPq and FAPESP, Brazil; NSERC, NRC and CFI, Canada; CERN; CONICYT, Chile; CAS, MOST and NSFC, China; COLCIENCIAS, Colombia; MSMT CR, MPO CR and VSC CR, Czech Republic; DNRF, DNSRC and Lundbeck Foundation, Denmark; ARTEMIS, European Union; IN2P3-CNRS, CEA-DSM/IRFU, France; GNAS, Georgia; BMBF, DFG, HGF, MPG and AvH Foundation, Germany; GSRT, Greece; ISF, MINERVA, GIF, DIP and Benoziyo Center, Israel; INFN, Italy; MEXT and JSPS, Japan; CNRST, Morocco; FOM and NWO, Netherlands; RCN, Norway;  MNiSW, Poland; GRICES and FCT, Portugal;  MERYS (MECTS), Romania;  MES of Russia and ROSATOM, Russian Federation; JINR; MSTD, Serbia; MSSR, Slovakia; ARRS and MVZT, Slovenia; DST/NRF, South Africa; MICINN, Spain; SRC and Wallenberg Foundation, Sweden; SER,  SNSF and Cantons of Bern and Geneva, Switzerland;  NSC, Taiwan; TAEK, Turkey; STFC, the Royal Society and Leverhulme Trust, United Kingdom; DOE and NSF, United States of America.  

The crucial computing support from all WLCG partners is acknowledged gratefully, in particular from CERN and the ATLAS Tier-1 facilities at TRIUMF (Canada), NDGF (Denmark, Norway, Sweden), CC-IN2P3 (France), KIT/GridKA (Germany), INFN-CNAF (Italy), NL-T1 (Netherlands), PIC (Spain), ASGC (Taiwan), RAL (UK) and BNL (USA) and in the Tier-2 facilities worldwide.

%--------------------------------- End of document ----------------------------------

\section*{References}

\bibliography{DijetMassAngle}

% ATLAS authorlist

\begin{flushleft}
{\Large The ATLAS Collaboration}

\bigskip

G.~Aad$^{\rm 48}$,
B.~Abbott$^{\rm 111}$,
J.~Abdallah$^{\rm 11}$,
A.A.~Abdelalim$^{\rm 49}$,
A.~Abdesselam$^{\rm 118}$,
O.~Abdinov$^{\rm 10}$,
B.~Abi$^{\rm 112}$,
M.~Abolins$^{\rm 88}$,
H.~Abramowicz$^{\rm 153}$,
H.~Abreu$^{\rm 115}$
% ...
\bigskip

% ...
$^{10}$ Institute of Physics, Azerbaijan Academy of Sciences, Baku, Azerbaijan\\
$^{11}$ Institut de F\'isica d'Altes Energies and Universitat Aut\`onoma  de Barcelona and ICREA, Barcelona, Spain\\
% ...
$^{48}$ Fakult\"{a}t f\"{u}r Mathematik und Physik, Albert-Ludwigs-Universit\"{a}t, Freiburg i.Br., Germany\\
$^{49}$ Section de Physique, Universit\'e de Gen\`eve, Geneva, Switzerland\\
% ...
$^{88}$ Department of Physics and Astronomy, Michigan State University, East Lansing MI, United States of America\\
% ...
$^{111}$ Homer L. Dodge Department of Physics and Astronomy, University of Oklahoma, Norman OK, United States of America\\
$^{112}$ Department of Physics, Oklahoma State University, Stillwater OK, United States of America\\
% ...
$^{115}$ LAL, Univ. Paris-Sud and CNRS/IN2P3, Orsay, France\\
% ...
$^{118}$ Department of Physics, Oxford University, Oxford, United Kingdom\\
% ...
$^{153}$ Raymond and Beverly Sackler School of Physics and Astronomy, Tel Aviv University, Tel Aviv, Israel\\
% ...
\end{flushleft}

\end{document}